\documentclass[seceq]{ptptex}

\usepackage{graphicx}

\title{
Light Four-quark States and New Observations by BES%
}

\author{
Ailin \textsc{Zhang}$^{1,}$\footnote{ e-mail address:
zhangal@staff.shu.edu.cn}, Tao \textsc{Huang}$^{2,}$\footnote{
e-mail address: huangtao@mail.ihep.ac.cn} and Tom
\textsc{Steele}$^{3,}$\footnote{ e-mail address:
tom.steele@usask.ca} }

\inst{
$^1$ Department of Physics, Shanghai University, Shanghai
200444, China\\
$^2$ Institute of High Energy Physics, CAS, P. O. Box 918 (4), Beijing 100049, China\\
$^{3}$ Department of Physics and Engineering Physics, University of
Saskatchewan, Saskatoon, SK S7N 5E2, Canada }

\abst{
Four-quark states are discussed within the constituent quark model.
Incompleteness of existed studies of four-quark state with QCD sum
rule is analyzed. The masses of diquark cluster were determined by
QCD sum rules, and light four-quark states masses were obtained in
terms of the diquark. The four-quark state possibility of the newly
observed near-threshold $p\bar p$ enhancement, $X(1835)$, $X(1812)$
and $X(1576)$ by BES is discussed.}

\begin{document}

\maketitle

\section{Four-quark states in the constituent quark model}

After Gell-Mann~\cite{gellmann} first conjectured the existence of
multi-quark states, four-quark states were predicted to exist in a
consistent description of the hadron scattering
amplitudes~\cite{rosner}. In the constituent quark model, four-quark
states may consist of a $qq$ diquark and a $\bar q\bar q$
anti-diquark (tetraquark), or of two $q\bar q$ clusters due to the
spatial extension of the quarks. Study of four-quark states was
extensively performed in many constituent quark models in late
1970's and revived with recent experimental~\cite{nakano} and
theoretical~\cite{wilczek} developments. Four-quark states were
initially regarded as hadron molecules and charmonium
atoms~\cite{okun}. Subsequently, it was studied in MIT bag
model~\cite{jaffe}, color junction model~\cite{chan}, potential
model~\cite{isgur,torn,glozman,brink}, relativistic quark
model~\cite{ebert}, effective Lagrangian~\cite{fariborz}, QCD sum
rules~\cite{sum-rule1,sum-rule2,sum-rule3,sum-rule4,sum-rule5} and
other models~\cite{others}. A review on exotica could be found in
Ref.~\cite{jaffe2}.

A four-quark state is a many-body system, in which quarks have many
degrees of freedom such as color, flavor and spin etc. The
correlation of these degrees of freedom among quarks is complicated,
and so far the exact correlation is not clear. According to
Ref.~\cite{zhang}, both the $(qq)(\bar q\bar q)$ and the $(q\bar
q)(q\bar q)$ four-quark state may result in the same color, flavor
and spin representation. $(qq)(\bar q\bar q)$ may mix with $(q\bar
q)(q\bar q)$. Furthermore, flavor Crypto-exotic four-quark states
may mix with normal $q\bar q$ mesons. That is to say, intrinsic
color, flavor configurations could not be distinguished without the
establishment of some special observable. Unfortunately, no such an
observable has been definitely constructed, and experimentally
observed meson may be $|meson>=|q\bar q>+|(qq)(\bar q\bar
q)>+|(q\bar q)(q\bar q)>+\cdots$

Of course, the correlation of color, flavor and spin in hadrons is
inter-related since their total correlation has to obey various
symmetry constraints. In the tetraquark $(qq)(\bar q\bar q)$ ,
according to Ref.~\cite{jaffe2}, two quarks correlate
antisymmetrically in color, flavor and spin, separately, and the
diquark is in a ``good" diquark configuration $|qq,\bar 3_F,\bar
3_C,0\rangle$. The tetraquark state consisting of ``good" diquark
may be the most suitable object to study the behaviour of quark
confinement~\cite{zhang} though quark dynamics in four-quark state
is still not clear.

\section{SVZ sum rules and four-quark state}

QCD sum rules~\cite{svz} are a semi-phenomenological nonperturbative
method of relating fundamental parameters of QCD Lagrangian and
vacuum to parameters of hadrons. In terms of suitable interpolating
currents, correlators are constructed. On one hand, the correlator
can be expressed with parameters of QCD and vacuum through the
operator product expansion (OPE). On the other hand, the correlator
can be expressed with parameters of hadrons through the dispersion
relation.

To perform the analysis of a sum rule, the choice of a suitable
interpolating current is important. In the study of four-quark
states, different currents were used. In Ref.~\cite{sum-rule1}, a
$(q\bar q)(q\bar q)$ current was used; In Ref.~\cite{sum-rule2},
both $(q\bar q)^2$ and $(qq)(\bar q\bar q)$ currents were analyzed;
In Ref.~\cite{sum-rule3}, a $(cq)(\bar q\bar q)$ current was used;
In Ref.~\cite{sum-rule4}, a $(cu)(\bar s\bar u)$ current was used;
In Ref.~\cite{sum-rule5}, a $(ud)(\bar s\bar s)$ current was used.
These investigations are interesting and may give some hints to our
understanding of four-quark states. However, they are not complete.

First, some of the conclusions related to the diquark concept based
on these studies are not definitive. Diquarks may be the reality in
constituent quark models, but a diquark interpretation is not
meaningful in the framework of QCD sum rules. In view of the sum
rule approach, the constituent quark structures of multi-quark
states are difficult to detect through the couplings of
interpolating currents to hadrons~\cite{discussion}.

This point is much more easy to be realized in other ways. The Fierz
transformation will turn the $(\bar qq)(\bar qq)$ current into the
$(\bar q\bar q)(qq)$ current and vice-versa, and these two kinds of
currents can mix with each other under renormalization. Therefore,
it may not be meaningful to talk about diquarks in the framework of
sum rules based on local interpolating currents.

Furthermore, correlators of tetraquark/four-quark interpolating
currents have only been calculated to leading order in $\alpha_s$~
\cite{sum-rule1,sum-rule2,sum-rule3,sum-rule4,sum-rule5}. To make
the sum rule analyses reliable and predictable, suitable
interpolating currents (for example, mixed currents ($(qq)(\bar
q\bar q)+(q\bar q)(q\bar q)$)) and $\alpha_s$ corrections (involving
four-loop calculations and the renormalization of the dimension-six
composite operators) are required.

It may be possible to proceed with the study of four-quark states in
another way. We can determine the mass of a diquark cluster via QCD
sum rules, and subsequently construct four-quark states in terms of
these diquark constituents.

\section{Light tetraquark state and new observations by BES}
We first describe the results of an updated sum-rule study of the
mass of $J^P=0^+$ ``good" diquark. The flavor $(sq)$ ``good" diquark
current was taken as that given in Ref.~\cite{dosch,jamin}, and
with the input parameters~\cite{zhang}, the most ``suitable"
$m_{qq}$ and $m_{sq}$ are determined: $m_{qq}\sim
400$~MeV($s_0=1.2~GeV^2$) and $m_{sq}\sim 460$~MeV($s_0=1.2~GeV^2$).

Once the determined masses are regarded as the constituent masses of
diquark, it is possible to construct a four-quark provided that the
quark dynamics in hadrons is known. Here hadron masses were obtained
from constituent quarks as in Refs.~\cite{jaffe,others}.

The mass of the $L=0$ tetraquark state is
\begin{eqnarray*}
M\approx 2m_{[qq]}-3(\kappa_{qq})_{\bar 3},
\end{eqnarray*}
with $(\kappa_{qq})_{\bar 3}=103$ MeV, $(\kappa_{sq})_{\bar 3}=64$
MeV. The masses of $0^{++}$ tetraquark states $[\bar q\bar q][qq]$,
$[\bar q\bar q][sq]$ ($[\bar s\bar q][qq]$) and $[\bar s\bar q][sq]$
are therefore found to be $\sim 490$ MeV, $\sim 610$ MeV and $\sim
730$ MeV, respectively. Taking into account the decay features, in
the approximation we used, it is reasonable to identify $f_0(600)$
(or $\sigma$), $f_0(980)$, $a_0(980)$ and the unconfirmed
$\kappa(800)$ as the $0^{++}$ light tetraquark states.

Similarly, the mass of the $1^{--}~(L=1)$ and the $2^{++}~(L=2)$
tetraquark state is
\begin{eqnarray}
M_{4q}=m_d+m_{\bar d}+B_{d\bar d}{L(L+1)\over 2}.
\end{eqnarray}
In the following, $B_{d\bar d}$ is denoted as $B^\prime_q$,
$B^\prime_{1s}$ and $B^\prime_{2s}$($B^\prime_q>
B^\prime_{1s}>B^\prime_{2s}$~\cite{others}) with zero, one and two
strange quarks, respectively. Masses of the $1^{--}$ orbital excited
$[\bar q\bar q][qq]$, $[\bar s\bar q][qq]$ and $[\bar s\bar q][sq]$
are respectively determined by $\sim 490+B^\prime_q$~MeV, $\sim
610+B^\prime_{1s}$~MeV and $\sim 730+B^\prime_{2s}$~MeV. Theoretical
estimates of the masses of these tetraquark states are not given for
the sensitivity of $B_{d\bar d}$ to $\Lambda_{QCD}$.

So far, no four-quark state has been confirmed experimentally. In
the low energy region, there are several four-quark candidates in
experiments, including $f_0(600)$ (or $\sigma$), $f_0(980)$,
$a_0(980)$ and the unconfirmed $\kappa(800)$. They have a long
history of being interpreted as four-quark states \cite{jaffe,pdg}.

In addition to these candidates, some recent observations by BES
collaboration such as the near-threshold $p\bar p$
enhancement~\cite{bes1}, $X(1835)$~\cite{bes2}, $X(1812)$
~\cite{bes3} and $X(1576)$~\cite{bes4} were interpreted as
four-quark candidates~\cite{roy,li,shen,lipkin,ding,zhang}.

In our analyses, the $p\bar p$ enhancements, $X(1835)$ and $X(1812)$
are unlikely to be tetraquark states. $X(1576)$ is likely to be the
$1^{--}$ tetraquark state, which may be the $(\bar s\bar q)(sq)$
``exotic" orbital excited tetraquark state~\cite{zhang}. It may be
the first orbital excitation of $a_0(980)$ if its isospin is $I=1$,
and it may be the first orbital excitation of $f_0(980)$ if its
isospin is $I=0$. If this suggestion is true, other $1^{--}$ orbital
excited tetraquark states corresponding to $(\bar q\bar q)(qq)$
($\sim 1400$ MeV) and $(\bar s\bar q)(qq)$ ($\sim 1500$ MeV) are
expected.

\section{Conclusions and discussions}

Incompleteness of previous study of four-quark states with QCD sum
rules have been discussed. The masses of diquarks have been
determined through an updated QCD sum rule analysis, and the
resulting masses of some tetraquark states have been constructed in
the constituent quark model.

Based on these analyses, it is reasonable to identify $f_0(600)$ (or
$\sigma$), $f_0(980)$, $a_0(980)$ and the unconfirmed $\kappa(800)$
as the $0^{++}$ light tetraquark states. The new observations by BES
are unlike to be the light four-quark states except that $X(1576)$
may be an "exotic" first orbital excited $(sq)(\bar s\bar q)$
tetraquark state.

\section*{Acknowledgements}
A. Zhang thanks the Yukawa Institute for Theoretical Physics at
Kyoto University, where this work was completed during the YKIS2006
on ``New Frontiers on QCD". A. Zhang thanks A. Hosaka (RCNP at Osaka
University) very much for his hospitality. Tao Huang is supported in
part by National Natural Science Foundation of China. T.\ Steele is
supported in part by NSERC (Natural Sciences and Engineering and
Research Council of Canada).


\end{document}